\documentclass[12pt]{article}
\usepackage{cite}
\usepackage{hyperref}
\usepackage{textcomp}
\usepackage{amsmath,amssymb,amsfonts}
\usepackage[margin=1in]{geometry}
\usepackage{booktabs}
\usepackage{array}
\usepackage{amsthm}
\usepackage{graphicx}
\newtheorem{theorem}{Theorem}

\title{Evolutionary Dynamics of Acid Resistance in Tumors: A Mathematical Model}
\author{Prithvi Anickode \\
{\normalsize Arizona State University, \quad{panickod@asu.edu}}\\and\\ Fabio Augusto Milner \\
{\normalsize Arizona State University, \quad{fmilner@asu.edu}}}

\date{\today}

\begin{document}

\maketitle

\begin{abstract}
Acidosis in tumors arises from reprogrammed metabolism and compromised vasculature, creating a harsh, acidic microenvironment that drives the evolutionary selection of acid-resistant cell phenotypes. A mathematical model is proposed to integrate phenotypic evolution, microenvironmental acidification, and tumor density dynamics. Three key mechanisms are incorporated in it: frequency-dependent selection favoring acid-resistant cells below a critical pH, stress-induced phenotypic switching, and a positive feedback loop where resistant cells produce excess acid that intensifies selection pressure. The well-posedness of the model is established. Through numerical simulations across biologically relevant parameter regimes, we identify two therapeutically targetable parameters: baseline acid clearance rate (a proxy for vascular perfusion) and a protection factor (representing acid-resistance "machinery" effectiveness) as critical bifurcation parameters for resistance evolution. The model exhibits qualitatively distinct dynamics depending on phenotypic plasticity levels. In low-plasticity tumors, both parameters exhibit sharp bifurcations with strong parameter interactions: clearance and protection effects are context-dependent, with therapeutic interventions effective only within specific parameter ranges. In high-plasticity tumors, both parameters produce continuous, monotonic responses with independent, additive effects. These regime-dependent dynamics suggest that treatment strategies should adapt to tumor plasticity: in the formar, targeting perfusion alone is typically sufficient, though sequential therapy may be required if the perfusion rate approaches or exceeds the bifurcation threshold, whereas in the latter treatment might benefit from combination therapies addressing both parameters simultaneously.\{These findings suggest that a low-dimensional model can identify therapeutically targetable parameters governing resistance evolution, suggesting interventions that may prevent or reverse the harmful effect of acid-resistant phenotypes, which are associated with chemotherapy failure, immune evasion, and metastatic progression.
\end{abstract}

\medskip

\textbf{Dynamical system model, tumor acidosis, parenchyma cells competition}

\section{Introduction}

Tumors thrive in a complex microenvironment, distinguishing them from regular tissue through multiple physiological changes. One of the most common features is extracellular acidosis, as tumor pH values usually range from 6.5 to 7.0, while normal tissue maintains a pH of 7.4 \cite{anemone2019imaging}. This phenomenon is the result of two related processes: increased glycolytic metabolism in cancer cells, which is known as the Warburg effect \cite{vanderheiden2009warburg}, and irregular tumor vasculature that inhibits the clearance of waste products, including protons (H$^+$ ions) \cite{dvorak2009blood}. \ {This acid-resistant phenotype has been documented across multiple cancer types, including colorectal cancer, where acidosis promotes DNA damage responses and genomic instability \cite{aubert2024acidosis}, and lung cancer, where it facilitates metastatic colonization through extracellular matrix remodeling \cite{shie2023acidosis}.} \ {Clinically, acidosis-driven resistance has been implicated in treatment failure across multiple modalities: acidic extracellular pH reduces uptake of weakly basic chemotherapeutics such as doxorubicin \cite{raghunand1999ph}, impairs T-cell cytotoxicity and checkpoint inhibitor efficacy \cite{pilon2016neutralization}, and promotes survival of radioresistant cell populations \cite{corbet2017acidosis}.}

The acidic tumor microenvironment is not just a byproduct of cancer metabolism; it plays a key role in tumor development. Acidic conditions promote invasion and metastasis by activating matrix-degrading enzymes, facilitating stromal remodeling, and suppressing immune responses \cite{estrella2013acidity,boedtkjer2020acidic}. Furthermore, the harsh acidic environment triggers and selects for acid-resistant phenotypes \cite{greaves2012clonal}. These acid-resistant cells are typically characterized by upregulated proton exporters, including monocarboxylate transporters (MCTs), sodium-hydrogen exchangers (NHE1), and carbonic anhydrase IX (CAIX), allowing them to maintain a close to neutral intracellular pH while the extracellular environment remains acidic \cite{liao2024ph,pillai2019causes}.

Beyond natural selection, evidence suggests that plasticity, which is the ability of cells to reversibly change phenotypes in response to stimuli without altering their genetic code, is a significant driver of adaptation in cancer \cite{yuan2019phenotypic,meacham2013tumour}. Phenotypic switching can be accelerated by stress, causing rapid adaptation to environmental changes through epigenetic modifications, altered transcriptional programs, and metabolic reprogramming \cite{zhou2023tumor,hoek2014phenotype}. This phenomenon could play a significant role in the context of acidic stress, where cells are exposed to environmental pressure that favors quick adaptive responses \cite{boedtkjer2020acidic}.

\ {Several mathematical frameworks have addressed some components of this problem. Game-theoretic models have captured frequency-dependent selection among tumor phenotypes: Archetti \cite{archetti2014warburg} framed glycolysis as a collective action problem where acid production serves as a public good, while Kaznatcheev et al.\ \cite{kaznatcheev2017cancer} extended this to a “double goods'' game incorporating both acidification and vascularization, demonstrating how polyclonal equilibria emerge from replicator dynamics. More recently, Fiandaca et al.\ \cite{fiandaca2021hypoxia} developed a partial integro-differential equation model coupling continuous phenotypic evolution with oxygen and lactate dynamics, showing how spatial gradients drive selection for acid-resistant traits. However, these models treat phenotypes as either fixed strategies or continuously evolving traits, omitting stress-induced phenotypic switching, the rapid, reversible transitions triggered by environmental stress. Furthermore, none incorporate a positive feedback loop where acid-resistant cells, through upregulated glycolysis associated with proton export machinery, produce excess acid that intensifies the very selection pressure favoring resistance. Our model addresses this gap by integrating frequency-dependent selection, stress-induced plasticity, and resistance-driven acidification within a unified ODE framework.

In this work, we develop a mathematical model that integrates these mechanisms to investigate how two key parameters, baseline acid clearance and acid-resistance ``machinery'' effectiveness, affect resistance evolution. Through bifurcation analysis across different levels of phenotypic plasticity, we identify regime-dependent therapeutic strategies and establish plasticity as a critical determinant of treatment response in acidic tumor microenvironments. \ {Specifically, we ask: can a low-dimensional ODE model identify therapeutically targetable parameters that govern resistance evolution, and thereby suggest interventions capable of preventing or reversing the tumor phenotypes associated with treatment failure, immune evasion, and metastatic progression?}

\section{Model Formulation}

\subsection{State Variables}

The model is based on three primary state variables that capture the evolutionary, micro-environmental, and population-level dynamics of the tumor system. These are defined as follows:

\begin{align}
x(t) &= \text{Fraction of acid-resistant cells} \in [0,1], \\
h(t) &= \text{Normalized proton concentration} = \frac{[\text{H}^+]}{[\text{H}^+]_{\text{physiological}}}, \\
n(t) &= \text{Normalized tumor density} = \frac{N(t)}{K},
\end{align}
where $[\text{H}^+]_{\text{physiological}} = 10^{-7.4}\ \text{M}$ corresponds to a physiological pH of 7.4. Here, $K$ denotes the carrying capacity of the environment, and $N(t)$ represents the total tumor cell population at time $t$. Note that, because of our chosen binary phenotypical structure, $1-x$ represents the fraction of acid-sensitive tumor cells.

\vspace{1em}

\noindent
\textbf{Interpretation of the pH normalization.}  
The variable $h(t)$ serves as a dimensionless measure of extracellular acidity, normalized relative to physiological proton concentration. When $h = 1$, the local environment corresponds to a normal physiological pH of 7.4. Values of $h$ greater than one indicate an increasingly acidic microenvironment (lower pH), while smaller values correspond to alkalosis, though this is rarely observed in tumors. The relationship between $h$ and pH can be expressed as
\[
\text{pH} = 7.4 - \log_{10}(h).
\]

Clinically relevant tumor pH values typically fall below physiological levels, reflecting varying degrees of acidosis. For instance, a mild decrease in pH to 7.2 corresponds to $h \approx 1.6$, moderate acidosis at pH 7.0 yields $h \approx 2.5$, and more severe acidification at pH 6.8 and 6.5 correspond to $h \approx 4.0$ and $h \approx 7.9$, respectively. These values delineate the physiological range over which the tumor microenvironment exerts selective pressure favoring acid-resistant phenotypes. \ {While $h$ is mathematically defined for all positive values (as established in Theorem 1), biologically relevant tumor microenvironments constrain $h$ to approximately $[1, 8]$, corresponding to the pH range $[6.5, 7.4]$ \cite{anemone2019imaging}. Values beyond this range represent extreme conditions: $h < 1$ corresponds to alkalosis (pH $> 7.4$), rarely observed in solid tumors, while $h > 8$ (pH $< 6.5$) induces extensive cell death that naturally limits further acidification \cite{park2011acidic,worsley2022inducing}. Rather than explicitly modeling extracellular buffering capacity, our model incorporates implicit saturation mechanisms that prevent unbounded acid accumulation: the glycolytic inhibition factor $G(h)$ reduces acid production as $h$ increases, while the Hill function $d_S(h)$ causes population decline through acid-induced mortality. These mechanisms, combined with density-dependent clearance $\Gamma(n)$, ensure $h$ remains bounded within the physiologically relevant range in all simulations.}

\subsection{Governing Equations}

The model is given by the following 3-dimensional system of ODEs:

\begin{equation}
\label{eq:model}
\boxed{
\begin{aligned}
\frac{dx}{dt} &= x(1-x)S(h) + \mu(h)(1-x) - \nu x, \\[8pt]
\frac{dh}{dt} &= \alpha n(1 + \beta x) \cdot G(h) - \Gamma(n)(h - 1), \\[8pt]
\frac{dn}{dt} &= rn(1-n) - D(h,x)n,
\end{aligned}
}
\end{equation}
where the density-dependent rates and fractions are modeled as follows:
\begin{align}
S(h) &= \frac{s_0}{1 + e^{-\lambda(h - h_c)}} - c + \varphi d_S(h), && \text{(unit net selection differential rate)} \label{eq:selection}\\
d_S(h) &= \frac{d_{\max} h^m}{h_{50}^m + h^m}, && \text{(unit acid-induced death rate)} \label{eq:death}\\
\mu(h) &= \mu_0 h^p, && \text{(unit stress-induced switching rate)} \label{eq:switching}\\
G(h) &= \frac{K_g}{K_g + h}, && \text{(reduction factor from glycolytic inhibition)} \label{eq:glycolysis}\\
\Gamma(n) &= \frac{\gamma_0}{1 + \eta n}, && \text{(unit acid clearance rate)} \label{eq:clearance}\\
D(h,x) &= (1 - \varphi x)d_S(h). && \text{(unit population death rate)} \label{eq:avgdeath}
\end{align}

\subsection{Biological Interpretation}
\subsubsection{Model Scope}
{In this model we assume spatial homogeneity (well-mixed system) and represent 
resistance as a binary phenotype, rather than in a continuous spectrum. Parameters are treated as constant, though they may vary with tumor stage and micro-environmental context.}

\subsubsection{Dynamics of Acid Resistant Cell Population}

The resistant cell-population fraction, $x$, evolves through three processes:
\paragraph{1. Frequency-dependent selection, $x(1-x)S(h)$:} The factor $x(1-x)$ represents the well known replicator dynamic structure from evolutionary game theory \cite{salvioli2021contribution}, where selection is strongest when both phenotypes coexist. The unit net selection differential rate, $S(h)$, incorporates three drivers for selection:

\begin{enumerate}
    \item[a.] \textbf{pH-dependent advantage}: A logistic function that allows for resistant cells to only gain advantage in acidic conditions when $h$ exceeds a critical value $h_c\approx 3.5$ (corresponding to pH = 6.9, see the discussion around \eqref{eq:hc}). The steepness parameter, $\lambda$,  controls the sharpness of the selection transition.
    
    \item[b.] \textbf{Metabolic cost}, $c$: Maintaining acid-resistance machinery (MCT1/4, NHE1, CAIX) requires ATP expenditure  \cite{liao2024ph}, causing a fitness disadvantage for resistance.
    
    \item[c.] \textbf{Survival advantage, $\varphi d_S(h)$:} Resistant cells experience reduced acid-induced mortality, and this term represents the fitness differential due to differences in mortality. 
\end{enumerate}

\paragraph{2. Stress-induced phenotypic switching, $\mu(h)(1-x)$:}
Acidic stress promotes conversion through ROS damage, epigenetic changes, and stress pathways \cite{yuan2019phenotypic}. The functional form $\mu(h) = \mu_0 h^p$, with $p > 1$, models nonlinear responses to increasing acidity.

\paragraph{3. Phenotypic reversion, $\nu x$:}
Resistant cells may revert to sensitive through epigenetic modification loss or metabolic reprogramming \cite{yuan2019phenotypic}.

\subsubsection{Micro-environmental Acid Dynamics}

\paragraph{Acid production rate: $\alpha n(1 + \beta x) \cdot G(h)$}

Tumor cells produce lactic acid through aerobic glycolysis \cite{vanderheiden2009warburg}. 
Resistant cells expressing acid-resistance machinery show enhanced acid production capacity, with experimental evidence showing increases in lactate production under acidic stress conditions \cite{jamali2015hypoxia}. The factor $(1 + \beta x)$ captures this enhanced 
production, where $\beta$ quantifies the relative increase in acid output by resistant cells.

The factor $G(h) = \frac{K_g}{K_g + h}$ models product inhibition of glycolytic enzymes 
at severe acidosis \cite{england2014ph}, where $K_g$ indicates the normalized proton concentration at which glycolytic activity is reduced to 50\% of its 
maximum. As $h \to \infty$ (severe acidosis), $G(h) \to 0$, effectively shutting down glycolysis, while at physiological pH ($h \approx 1$), $G(h) \approx K_g/(K_g+1)$ allows near-maximal acid production.

\paragraph{Acid clearance rate: $\Gamma(n)(h - 1)$}

Clearance decreases with increasing tumor density due to vascular compression \cite{dvorak2009blood}. The factor $(h-1)$ drives pH toward physiological levels when clearance is effective.

\subsubsection{Population Dynamics}

We modeled the dynamics of tumor density through logistic growth, $rn(1-n)$, and pH-dependent mortality $D(h,x)$ using Hill kinetics. The unit death rate is reduced for resistant cells by a factor $(1-\varphi x)$, reflecting their enhanced survival under acidic stress.

\subsection{Mathematical Well-Posedness}

Before analyzing the biological implications of the model, we establish that it is mathematically well-posed and preserves biological constraints.

\begin{theorem}[Existence, Uniqueness, and Invariance]
For any initial condition $(x_0, h_0, n_0) \in [0,1] \times \mathbb{R}_{>0} \times [0,1]$ and positive parameter values, system \eqref{eq:model} admits a unique solution $(x(t), h(t), n(t))$ defined for all $t \geq 0$ that remains in the biologically relevant domain $\mathcal{D} = [0,1] \times \mathbb{R}_{>0} \times [0,1]$.
\end{theorem}

\begin{proof}

\textbf{Part 1: Local Existence and Uniqueness}

The right-hand side functions of system \eqref{eq:model} are compositions of smooth functions. Specifically:
\begin{itemize}
    \item $S(h)$ is $C^\infty$ on $(0,\infty)$ (exponential and rational functions)
    \item $\mu(h) = \mu_0 h^p$ is $C^\infty$ on $(0,\infty)$
    \item $d_S(h) = d_{\max}h^m/(h_{50}^m + h^m)$ is $C^\infty$ on $(0,\infty)$
    \item $G(h) = K_g/(K_g + h)$ is $C^\infty$ on $(0,\infty)$
    \item $\Gamma(n) = \gamma_0/(1 + \eta n)$ is $C^\infty$ on $[0,1]$
    \item $D(h,x) = (1-\varphi x)d_S(h)$ is $C^\infty$ on $[0,1] \times (0,\infty)$
\end{itemize}

Therefore, the vector field $F(x,h,n) = (\dot{x}, \dot{h}, \dot{n})^T$ is continuously differentiable on $\mathcal{D}$, hence locally Lipschitz continuous. By the Picard-Lindel\"of theorem \cite{perko2001differential}, there exists a unique local solution for any initial condition in $\mathcal{D}$. 

\textbf{Part 2: Invariance of $[0,1]$ for $x(t)$}

We show that $[0,1]$ is positively invariant for $x(t)$.

At $x = 0$:
\begin{equation}
\frac{dx}{dt}\bigg|_{x=0} = 0 \cdot (1-0) \cdot S(h) + \mu(h)(1-0) - \nu \cdot 0 = \mu(h) = \mu_0 h^p > 0
\end{equation}
for all $h > 0$, $\mu_0 > 0$, and $p > 0$. Thus, the vector field points to the interior of $\mathcal{D}$ at $x = 0$, and $x(t)$ cannot become negative.

At $x = 1$:
\begin{equation}
\frac{dx}{dt}\bigg|_{x=1} = 1 \cdot 0 \cdot S(h) + \mu(h) \cdot 0 - \nu = -\nu < 0
\end{equation}
since $\nu > 0$. Thus, the vector field points to the interior of $\mathcal{D}$ at $x = 1$, and $x(t)$ cannot exceed 1.

By the continuous dependence of solutions, $x(t) \in [0,1]$ for all $t \geq 0$ whenever $x_0 \in [0,1]$.

\textbf{Part 3: Positivity of $h(t)$}

We establish that $h(t) > 0$ for all $t \geq 0$ if $h_0 > 0$. The $h$-dynamics are given by:
\begin{equation}
\frac{dh}{dt} = \alpha n(1 + \beta x) G(h) - \Gamma(n)(h - 1)
\end{equation}

At $h = 0^+$ (0 approached from the right):
\begin{equation}
\frac{dh}{dt}\bigg|_{h \to 0^+} = \alpha n(1 + \beta x) \cdot \frac{K_g}{K_g + 0} - \Gamma(n)(0 - 1) = \alpha n(1 + \beta x) + \Gamma(n) > 0
\end{equation}
since $\alpha, \beta, K_g, n, x \geq 0$ and $\Gamma(n) = \gamma_0/(1 + \eta n) > 0$. The factor $\alpha n(1+\beta x) \geq 0$ in the production term and factor $\Gamma(n) > 0$ in the clearance term  ensure the right-hand side is strictly positive at $h = 0$.

Therefore, if $h(t)$ were to reach 0 from above, the derivative would be positive, forcing $h$ to increase. This prevents $h(t)$ from becoming non-positive, and thus $h(t) > 0$ for all $t \geq 0$ if $h_0 > 0$.

\textbf{Part 4: Invariance of $[0,1]$ for $n(t)$}

We show that $[0,1]$ is positively invariant for $n(t)$.

At $n = 0$:
\begin{equation}
\frac{dn}{dt}\bigg|_{n=0} = r \cdot 0 \cdot (1-0) - D(h,x) \cdot 0 = 0
\end{equation}
Thus, $n = 0$ is invariant. If the population is extinct ($n = 0$), it remains extinct.

At $n = 1$:
\begin{equation}
\frac{dn}{dt}\bigg|_{n=1} = r \cdot 1 \cdot (1-1) - D(h,x) \cdot 1 = -D(h,x) \leq 0
\end{equation}
where $D(h,x) = (1-\varphi x)d_S(h) \geq 0$ for all $(h,x) \in (0,\infty) \times [0,1]$ since $\varphi \in [0,1]$ and $d_S(h) \geq 0$. The inequality is strict (i.e., $dn/dt < 0$) when $h$ is sufficiently large that $d_S(h) > 0$, because $\varphi x < 1$. Thus, the vector field points to the interior of $\mathcal{D}$ at $n = 1$, preventing $N$ from exceeding carrying capacity.

For $n \in (0,1)$, the logistic growth structure ensures $n$ remains bounded within $(0,1]$ for any positive initial density $n_0 \in (0,1]$.

\textbf{Part 5: Global Existence}

Having established that $\mathcal{D} = [0,1] \times \mathbb{R}_{>0} \times [0,1]$ is positively invariant, solutions starting in $\mathcal{D}$ remain bounded for all $t \geq 0$. Bounded solutions cannot exhibit finite-time blow-up. Therefore, the local solution extends to a global solution defined for all $t \geq 0$.

Combining Parts 1-5, we conclude that system \eqref{eq:model} admits a unique global solution in $\mathcal{D}$ for all $t \geq 0$. \qedhere
\end{proof}

This result ensures that the model respects biological constraints and produces well-defined dynamics for any biologically reasonable initial condition.

\subsection{Equilibrium Points}
Let us denote $E=\left(x^*,h^*,n^*\right)$ a generic equilibrium point of the system \eqref{eq:model}. 

We see that the second and third equations in \eqref{eq:model} admit the trivial equilibrium $E_0=(x^*,1,0)$, where $x^*$ is determined by the first equation in \eqref{eq:model} with $h=1$ as:
\begin{equation}\label{eq:x*trivial}
x^*=\displaystyle\frac{\displaystyle\sqrt{\left(\frac{\mu_0+\nu}{S(1)}-1\right)^2+4\frac{\mu_0}{S(1)}}-\left(\frac{\mu_0+\nu}{S(1)}-1\right)}2,
\end{equation}
and it is easy to check that $0<x^*<1$ is thus uniquely  determined. This is the extinction equilibrium where the tumor is totally reabsorbed, as the pH approaches its physiological value and the fraction of resistant cells hovers around the value given in \eqref{eq:x*trivial}, while their number necessarily approaches 0.

\ {Note that, for $h\equiv1$ and $x\equiv x^*$, we can solve the equation for the total cell population density, $n$, in \eqref{eq:model} explicitly, using the notation $D^*=D(1,x^*)$:
\begin{equation}\label{eq:ntrivial}
n(t)=\displaystyle\frac{D^*-r}{\left(r+\frac{D^*-r}{n(0)}\right)e^{(D^*-r)T}-r},
\end{equation}
whereby 
\begin{equation}\label{limitn}
\displaystyle\lim_{t\to\infty}n(t)=\left\{
\begin{aligned}
    0,&\quad\mbox{if\quad}\frac r{D^*}<1,\\
    1-\frac1{D^*/r},&\quad\mbox{if\quad}\frac r{D^*}>1.
\end{aligned}
\right.
\end{equation}
We can see in \eqref{limitn} that 
\begin{equation}\label{R0}
\mathcal{R}_0=\displaystyle\frac r{D^*}
\end{equation}
is the basic reproduction number of the total cell population of the tumor at acidity level $h=1$ and fraction of resistant cells $x=x^*$.
Concerning the stability of $E_0$, we have the following result based on this basic reproduction number.}

\begin{theorem}
    The equilibrium $E_0$ is locally asymptotically stable if\/ $\mathcal{R}_0<1$ and it is unstable if\/ $\mathcal{R}_0>1$.
\end{theorem}
\begin{proof}
    The Jacobian matrix of the system \eqref{eq:model} at the equilibrium $E_0$ is given as
 \begin{equation}
    J(E_0)= \left(
 \begin{matrix}
        -\displaystyle\sqrt{(\mu_0+\nu-S(1))^2+4\mu_0S(1)}&\qquad ***&\qquad ***\\
        0&\qquad-\gamma_0&\qquad\alpha(1+\beta x^*)\displaystyle\frac{K_g}{K_g+1}\\
        0&\qquad0&\qquad r-D(1,x^*)
    \end{matrix}
    \right)
    \end{equation}
This matrix is upper triangular with the first two diagonal coefficients negative. The theorem follows from the observation that the last coefficient of the matrix is negative if, and only if, $\mathcal{R}_0<1$ and it is positive if, and only if, $\mathcal{R}_0>1$.
\end{proof}

\medskip

\ {This theorem suggests that $\mathcal{R}_0$ acts as a bifurcation parameter and, at $\mathcal{R}_0=1$ we have a forward bifurcation from the local stability of $E_0$ for $\mathcal{R}_0<1$ to its loss of stability for $\mathcal{R}_0>1$, when another (positive) equilibrium appears and inherits the local stability.
}

\medskip

The only equilibrium with $n^*=0$ is $E_0$, so that we now seek other steady-states with $n^*>0$. Note that the first equation in \eqref{eq:model} together with \eqref{eq:switching} imply that, if $x^*=0$, then $h^*=0$, which is impossible (this corresponds to infinite pH). Therefore, there is no equilibrium without a positive fraction of resistant cells.

However, an equilibrium without sensitive cells is possible: if $x^*=1$, then necessarily $\nu=0$, which would necessitate the removal of the assumption that resistant cells can revert to sensitive. In that case, the equilibrium 
\begin{equation}
    E_1=(1,h^*,n^*)
\end{equation}
is determined by the last two equations of \eqref{eq:model}. The last one, together with \eqref{eq:death} and \eqref{eq:avgdeath}, imply that $$
1-n^*=(1-\varphi)\;\displaystyle\frac{(h^*)^m}{(h_{50})^m+(h^*)^m}\;\frac{d_{\mathrm{max}}}r.
$$ 
For this equation to have positive solutions, we need $\mathcal{R}^*_0=\frac r{d_{\mathrm{max}}}>1$. In that case,
\begin{equation}\label{n*}
    n^*=1-(1-\varphi)\;\displaystyle\frac{(h^*)^m}{(h_{50})^m+(h^*)^m}\;\frac{d_{\mathrm{max}}}r\in(0,1).
\end{equation}
The second equation in \eqref{eq:model} implies that 
\begin{equation}\label{equil1}
\alpha(1+\beta)K_g\;n^*(1-n^*)=\gamma_0(h^*-1)(K_g+h^*),
\end{equation}
and we see that $n^*>0\iff h^*>1$. Substituting $n^*$ from \eqref{n*} into \eqref{equil1}, we obtain an equation for the equilibrium acidity level, $h^*$:
\begin{equation}\label{h*}
\alpha(1+\beta)K_g\left(1-(1-\varphi)\;\displaystyle\frac{(h^*)^m}{(h_{50})^m+(h^*)^m}\;\frac{d_{\mathrm{max}}}r\right)\left((1-\varphi)\;\displaystyle\frac{(h^*)^m}{(h_{50})^m+(h^*)^m}\;\frac{d_{\mathrm{max}}}r\right)=\gamma_0(h^*-1)(K_g+h^*).
\end{equation}
We are seeking the roots of a polynomial of degree $2m+2$, and we will use $m=2$ in our simulations. This means we need the roots of polynomial of degree 6, which cannot be expressed in closed form in terms of the coefficients. However, comparing the signs of the left-hand side and the right-hand side of \eqref{h*}, we see that any positive root must satisfy $h^*>1)$. Moreover, as $h$ increases from 1 to $+\infty$, the left-hand side of the equation is always positive and bounded, increasing to a maximum and then decreasing towards a horizontal asymptote, while the right-hand side grows monotonically from 0 to $+\infty$. Hence, there is a unique root $h^*>1$, and the corresponding $n^*$ given by \eqref{n*} lies in the interval $(0,1)$. 

We have established the following result.
\begin{theorem}
    Let $\mathcal{R}^*_0=\frac r{d_{\mathrm{max}}}>1$ and assume $\nu=0$. Then there is a unique positive equilibrium of \eqref{eq:model} with $x^*=1$, $E_1=(1,h^*,n^*)\in\mathcal{D}$.
\end{theorem}

We were unable to \ {prove} stability results for this equilibrium, and we were also unable to \ {find an explicit formula for} the positive equilibria with $\nu>0$, corresponding to $0<x^*<1$.
\subsection{Parameter Estimation}

Model parameters are estimated through a phenomenological approach, choosing values that produce biologically plausible dynamics while exploring qualitatively different model behaviors. Ranges are constrained by general biological knowledge and dimensional consistency, with specific values selected through numerical exploration to identify parameter regimes yielding distinct evolutionary outcomes.
\subsubsection{Population Dynamics Parameters}

\paragraph{Intrinsic growth rate ($r$)}

The normalized tumor density $n(t) = N(t)/K$ evolves according to logistic growth, where $r$ controls the rate of approach to carrying capacity:
\begin{equation}
r \in [0.01, 0.05] \text{ day}^{-1}
\end{equation}

This range spans slow-growing tumors reaching half-capacity in 140 days ($r = 0.01$) to aggressive tumors reaching half-capacity in 28 days ($r = 0.05$). Baseline $r = 0.03$ day$^{-1}$ represents moderately aggressive growth.

\paragraph{Acid-induced death parameters ($d_{\max}$, $h_{50}$, $m$)}

Experimental studies demonstrate that acidic extracellular pH induces apoptosis in cancer cells, with significant cell death observed at pH 6.5-6.8 within 24-48 hours \cite{park2011acidic,worsley2022inducing}. We adopt ranges that capture varying sensitivity to acidosis across tumor types:

\begin{align}
d_{\max} &\in [0.05, 0.15] \text{ day}^{-1} && \text{(maximum death rate)} \\
h_{50} &\in [3.5, 6.5] && \text{(half-maximal effect at pH 6.5-7.0)} \\
m &\in [2, 4] && \text{(Hill coefficient for cooperativity)}
\end{align}

Higher $h_{50}$ values indicate cells that maintain viability at more severe acidosis levels before death rates become substantial. Baseline values are $d_{\max} = 0.08$ day$^{-1}$, $h_{50} = 4.5$, and $m = 2.5$.

\subsubsection{Selection and Fitness Parameters}

\paragraph{Maximum selection advantage rate ($s_0$)}

The maximum selection advantage determines how strongly acidosis favors resistant phenotypes:
\begin{equation}
s_0 \in [0.10, 0.30] \text{ day}^{-1}
\end{equation}

These values enable selection to significantly influence evolutionary dynamics on timescales comparable to population growth. Baseline $s_0 = 0.20$ day$^{-1}$.

\paragraph{Metabolic cost ($c$)}

Maintaining proton exporters (MCT1/4, NHE1, CAIX) requires ATP expenditure, imposing a constitutive fitness cost:
\begin{equation}
c \in [0.01, 0.04] \text{ day}^{-1}
\end{equation}

This range reflects varying metabolic burdens depending on expression levels and activity of acid-resistance machinery. Baseline $c = 0.02$ day$^{-1}$.

\paragraph{Protection factor ($\varphi$)}

Represents fractional reduction in acid-induced mortality for resistant cells:
\begin{equation}
\varphi \in [0.60, 0.95] \quad \text{(60-95\% protection)}
\end{equation}

Higher values model cells with highly effective acid-resistance mechanisms that nearly eliminate acid-induced mortality. Baseline $\varphi = 0.75$.

\paragraph{Critical threshold ($h_c$)}

The critical threshold represents the normalized proton concentration at which acid-resistant phenotypes begin gaining selective advantage. Experimental evidence indicates normal cells show impaired function below pH 7.0 ($h \approx 2.5$), while acid-resistant cells expressing MCT1/4, NHE1, and CAIX maintain viability at lower pH \cite{liao2024ph,pillai2019causes}:
\begin{equation}\label{eq:hc}
h_c \in [2.5, 4.5] \quad \text{(pH 7.0-6.8)}
\end{equation}

Lower values indicate resistance becomes advantageous at milder acidosis, while higher values require more severe conditions. Baseline $h_c = 3.5$ (pH 6.9).

\paragraph{Transition steepness ($\lambda$)}

Controls sharpness of the sigmoidal selection transition:
\begin{equation}
\lambda \in [3, 6] \quad \text{(sharp to very steep transitions)}
\end{equation}

Larger values create more switch-like transitions between selection regimes. Baseline $\lambda = 4.5$.

\subsubsection{Phenotypic Switching Parameters}

\paragraph{Stress-induced switching ($\mu_0$, $p$)}

Phenotypic plasticity involves stochastic transitions reflecting epigenetic flexibility:
\begin{align}
\mu_0 &\in [5 \times 10^{-5}, 1 \times 10^{-3}] \text{ day}^{-1} && \text{(baseline unit switching rate)} \\
p &\in [1.5, 3] && \text{(stress response nonlinearity parameter)}
\end{align}

Lower $\mu_0$ values model cells with strong epigenetic barriers to switching, while higher values allow more rapid phenotypic adaptation.  The nonlinearity parameter of the stress response, $p$, controls sensitivity to acidic stress. Baseline values $\mu_0 = 2 \times 10^{-4}$ day$^{-1}$ and $p = 2.0$.

\paragraph{Reversion rate ($\nu$)}

Reversion from resistant to sensitive states:
\begin{equation}
\nu \in [1 \times 10^{-4}, 5 \times 10^{-4}] \text{ day}^{-1}
\end{equation}

Values are chosen to allow moderate reversibility, preventing complete fixation while maintaining evolutionary responsiveness. Baseline $\nu = 3 \times 10^{-4}$ day$^{-1}$.

\subsubsection{Acid Production and Clearance}

\paragraph{Basal unit acid production rate ($\alpha$)}

Enhanced glycolysis and lactate production in tumors occur on faster time-scales than population growth:
\begin{equation}
\alpha \in [0.07, 0.15] \text{ day}^{-1}
\end{equation}

These values ensure pH changes occur rapidly relative to population dynamics, enabling micro-environmental acidification to drive evolutionary selection. Baseline $\alpha = 0.10$ day$^{-1}$.

\paragraph{Percentual excess production by resistant cells ($\beta$)}

Acid-resistant phenotypes upregulate glycolysis to fuel proton export. Experimental measurements in breast cancer cells show that expression of acid-resistance machinery (CAIX) under acidic stress increases lactate production up to 60\% \cite{jamali2015hypoxia}:
\begin{equation}
\beta \in [0.35, 0.65]
\end{equation}

Baseline $\beta = 0.50$ represents a 50\% increase in acid production by resistant cells, consistent with experimental observations.

\paragraph{Glycolytic inhibition ($K_g$)}

Product inhibition factor of glycolytic enzymes, particularly phosphofructokinase, becomes significant at pathological pH \cite{england2014ph}:
\begin{equation}
K_g \in [2.0, 4.0]
\end{equation}

Lower values indicate earlier onset of glycolytic inhibition, while higher values allow glycolysis to persist under more severe acidosis. Baseline $K_g = 3.0$.

\paragraph{Clearance parameters ($\gamma_0$, $\eta$)}

Vascular clearance depends on perfusion, which degrades with tumor density due to vascular compression and irregular structure:
\begin{align}
\gamma_0 &\in [\ {0.01, 0.06}] \text{ day}^{-1} && \text{(maximal unit clearance rate)} \\
\eta &\in [0.15, 0.35] && \text{(perfusion reduction strength)}
\end{align}

These ranges reflect severely compromised vasculature characteristic of solid tumors, with clearance occurring much slower than acid production. Baseline values $\gamma_0 = 0.03$ day$^{-1}$ and $\eta = 0.25$.

\subsection{Baseline Parameter Set}

Table \ref{tab:baseline_params} summarizes baseline parameter values for numerical simulations. These represent a moderately aggressive tumor with compromised vascular clearance.

\begin{table}[htbp]
\centering
\caption{Model parameters: definitions, units, baseline values, and biologically plausible ranges}
\label{tab:baseline_params}
\footnotesize
\begin{tabular}{p{1.2cm}p{5.5cm}p{1.5cm}p{1.5cm}p{2cm}}
\toprule
\textbf{Symbol} & \textbf{Definition} & \textbf{Units} & \textbf{Baseline} & \textbf{Range} \\
\midrule
\multicolumn{5}{l}{\textit{Population Dynamics}} \\
\midrule
$r$ & Intrinsic unit tumor growth rate & day$^{-1}$ & 0.03 & [0.01, 0.05] \\
$d_{\max}$ & Maximal acid-induced unit death rate & day$^{-1}$ & 0.08 & [0.05, 0.15] \\
$h_{50}$ & Normalized proton concentration at half-maximal mortality & \ \ -- & 4.5 & [3.5, 6.5] \\
$m$ & Hill coefficient for acid-induced mortality cooperativity & \ \ -- & 2.5 & [2, 4] \\
\midrule
\multicolumn{5}{l}{\textit{Selection and Fitness}} \\
\midrule
$s_0$ & Maximum selection advantage rate of resistant cells & day$^{-1}$ & 0.20 & [0.10, 0.30] \\
$c$ & Unit metabolic cost of maintaining resistance machinery & day$^{-1}$ & 0.02 & [0.01, 0.04] \\
$\varphi$ & Protection factor (fractional reduction in unit death rate) & \ \ -- & 0.75 & [0.60, 0.95] \\
$h_c$ & Critical proton concentration threshold for selection & \ \ -- & 3.5 & [2.5, 4.5] \\
$\lambda$ & Steepness of selection transition & \ \ -- & 4.5 & [3, 6] \\
\midrule
\multicolumn{5}{l}{\textit{Phenotypic Switching}} \\
\midrule
$\mu_0$ & maximal stress-induced unit switching rate & day$^{-1}$ & $2 \times 10^{-4}$ & [$5 \times 10^{-5}$, $10^{-3}$] \\
$p$ & Nonlinearity exponent for stress response & \ \ -- & 2.0 & [1.5, 3] \\
$\nu$ & Unit reversion rate from resistant to sensitive & day$^{-1}$ & $3 \times 10^{-4}$ & [$10^{-4}$, $5 \times 10^{-4}$] \\
\midrule
\multicolumn{5}{l}{\textit{Acid Dynamics}} \\
\midrule
$\alpha$ & Basal acid production unit rate & day$^{-1}$ & 0.10 & [0.07, 0.15] \\
$\beta$ & Relative excess acid production by resistant cells & \ \ -- & 0.50 & [0.35, 0.65] \\
$K_g$ & Glycolytic inhibition constant & \ \ -- & 3.0 & [2.0, 4.0] \\
$\gamma_0$ & Maximal vascular clearance unit rate & day$^{-1}$ & 0.03 & [\ {0.01, 0.06}] \\
$\eta$ & Strength of density-dependent clearance reduction & \ \ -- & 0.25 & [0.15, 0.35] \\
\bottomrule
\end{tabular}
\end{table}
\section{Numerical Simulations}
We perform numerical simulations to investigate the long-term evolutionary dynamics of the model 
and identify the role of unit acid clearance rate ($\gamma_0$) and protection factor ($\varphi$) as critical control parameters. Due to the presence of transcendental equations when solving for equilibrium points, analytical characterization of bifurcations is intractable, which is why a computational approach was used.

\subsection{Methods}

All simulations were performed using Python 3.11 with the \texttt{scipy.integrate.odeint} function for numerical integration of system \eqref{eq:model}. We used adaptive time-stepping with relative tolerance $10^{-8}$ and absolute tolerance $10^{-10}$ to ensure accurate resolution of the coupled nonlinear dynamics.

\textbf{Time-series analysis:} For trajectory visualization, the system was integrated from $t = 0$ to $t = 800$ days from initial condition $(x_0, h_0, n_0) = (0.02, 1.0, 0.15)$, representing a small initial resistant subpopulation at physiological pH and low tumor density.

\textbf{Single-parameter bifurcation analysis:} To characterize equilibrium dependence on individual parameters, we systematically varied one parameter across its biologically plausible range (Table~\ref{tab:baseline_params}) while holding all others at baseline. At each parameter value, the system was integrated to $t = 2500$ days, and equilibrium resistant fraction $x^*$ was recorded after verifying steady-state convergence ($|dx/dt| < 10^{-6}$). 

\ {\textbf{Parameter Regimes}:} We explored two distinct parameter regimes, both biologically plausible but differing in phenotypic plasticity and resistance effectiveness. \ {The specific parameter values defining these regimes were chosen phenomenologically to produce qualitatively distinct dynamics; empirical quantification of stress-induced switching rates in tumor cells under acidic conditions remains an important area for future experimental work.}

\textbf{Regime A (Low Plasticity):} Characterized by low stress-induced switching rate ($\mu_0 = 5 \times 10^{-5}$ day$^{-1}$) and moderate protection ($\varphi = 0.70$), representing tumors with limited phenotypic plasticity. All other parameters set to baseline values found in table~\ref{tab:baseline_params}.

\textbf{Regime B (High Plasticity):} Characterized by high stress-induced switching rate ($\mu_0 = 8 \times 10^{-4}$ day$^{-1}$) and strong protection ($\varphi = 0.95$), representing highly adaptive tumors with robust acid-resistance machinery. All other parameters set to baseline values found in table~\ref{tab:baseline_params}.

\textbf{Two-parameter phase space mapping:} For Regime B, we computed equilibrium resistance on a 50 × 50 grid spanning clearance rate $\gamma_0 \in [0.018, 0.038]$ day$^{-1}$ and protection factor $\varphi \in [0.30, 0.98]$. At each grid point, the system was integrated to steady state using the method above.

\ {\textbf{Global Sensitivity Analysis:} To quantify parameter influence on equilibrium resistance while accounting for parameter interactions, we performed global sensitivity analysis using Latin Hypercube Sampling (LHS) with Partial Rank Correlation Coefficients (PRCC) \cite{marino2008methodology}. We generated $N = 1000$ samples across the 17-dimensional parameter space, with each parameter sampled uniformly from its biologically plausible range (Table~\ref{tab:baseline_params}). For each sample, the system was integrated to equilibrium and the resistant fraction $x^*$ recorded. PRCC values quantify the monotonic relationship between each parameter and $x^*$ while controlling for all other parameters.}


\subsection{Regime A: Sharp Bifurcations}

\subsubsection{Time-Series Dynamics}

In Regime A, the model exhibits threshold-dependent dynamics where a critical value of $\gamma_0$ determines qualitatively distinct evolutionary outcomes (Figure \ref{fig:regime_a}).

\textbf{Low clearance scenario ($\gamma_0 = 0.018$ day$^{-1}$):}
The tumor evolves toward a high-resistance state. Starting from 2\% resistant cells, the resistant fraction increases to $76.4\%$ over 800 days. During the same period, tumor pH equilibrates to 6.80, corresponding to severe acidosis ($h \approx 4.0$) and tumor density equilibrates at $76.4\%$, below carrying capacity, due to acid-induced mortality. The system reaches a stable equilibrium characterized by a predominantly resistant population maintaining a chronically acidic micro-environment.

\textbf{High clearance scenario ($\gamma_0 = 0.055$ day$^{-1}$):}
In contrast, enhanced vascular clearance leads to suppression of resistance evolution. The resistant fraction remains low throughout the simulation, equilibrating at approximately $5\%$. Tumor pH equilibrates at 7.09, representing only mild acidosis ($h \approx 1.9$). Tumor density equilibrates at $81.5\%$, closer to carrying capacity, due to reduced acid stress. The equilibrium corresponds to a predominantly sensitive population in a nearly physiological micro-environment.

\begin{figure}[htbp]
\centering
\includegraphics[width=\textwidth]{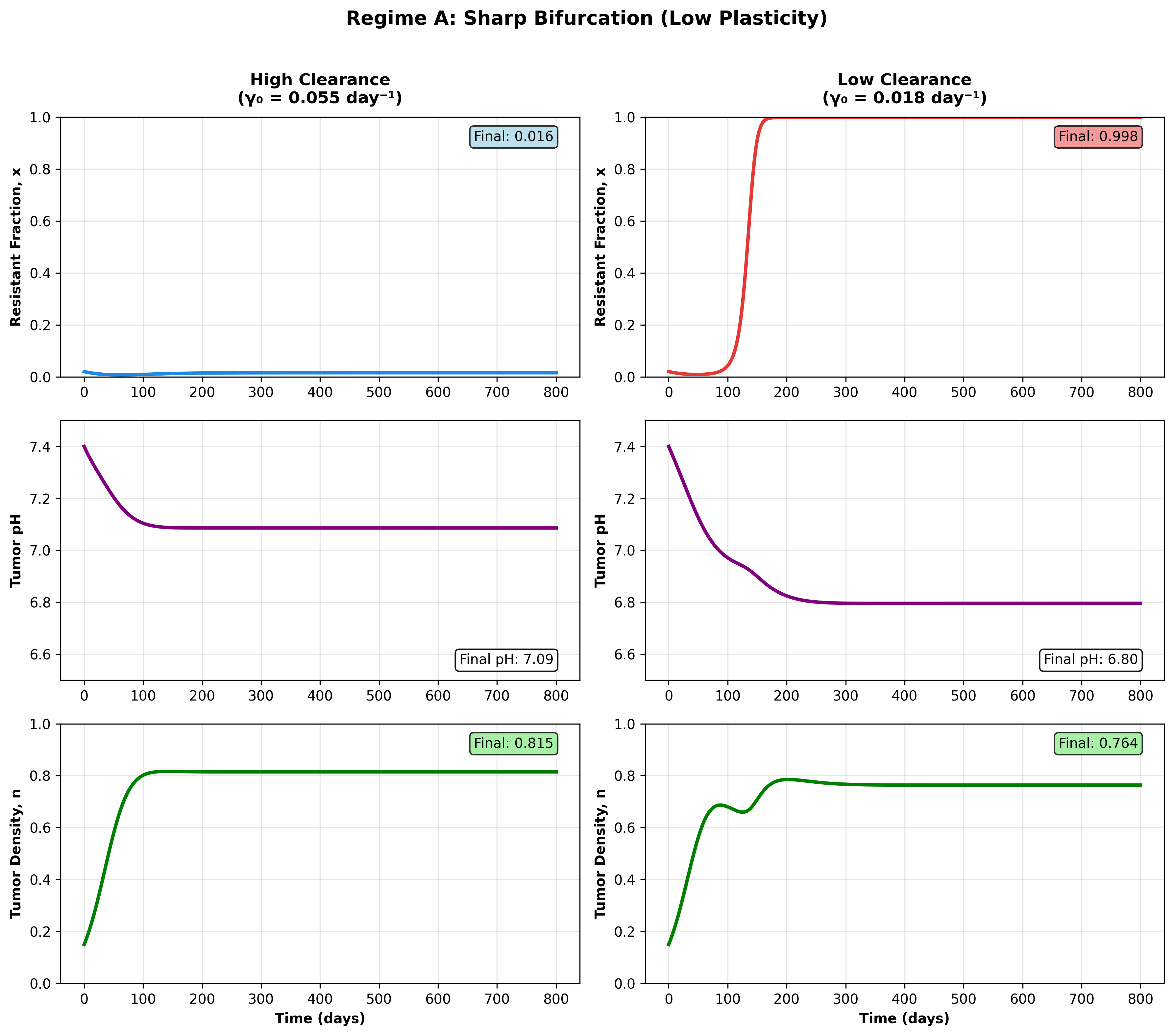}
\caption{\textbf{Regime A: Time-series dynamics in low-plasticity tumors.} Time evolution of resistant fraction, $x(t)$, tumor pH, and tumor density, $n(t)$, for low clearance ($\gamma_0 = 0.018$ day$^{-1}$, left) and high clearance ($\gamma_0 = 0.055$ day$^{-1}$, right). All other parameters identical. Initial conditions: $(x_0, h_0, n_0) = (0.02, 1.0, 0.15)$. Parameter values shown in table.}
\label{fig:regime_a}
\end{figure}

\subsubsection{Bifurcation with Respect to Clearance Rate}

To systematically characterize the dependence of equilibrium resistance on clearance rate, we computed equilibrium values of $x$ across a range of $\gamma_0$ values (Figure \ref{fig:bifurcation_a_gamma}). The bifurcation diagram reveals a sharp transition: below $\gamma_0 \approx 0.039$ day$^{-1}$, the system equilibrates to near-complete resistance, while above this value, it transitions to near-complete sensitivity. It is important to note that complete resistance and sensitivity cannot be achieved due to the model structure allowing constant switching and reversion. The sharp switch in equilibrium resistance indicates a bifurcation, where two equilibrium points, one corresponding to resistant dominance and the other to sensitive dominance, exchange stability at this critical value of $\gamma_0$.

\begin{figure}[htbp]
\centering
\includegraphics[width=0.8\textwidth]{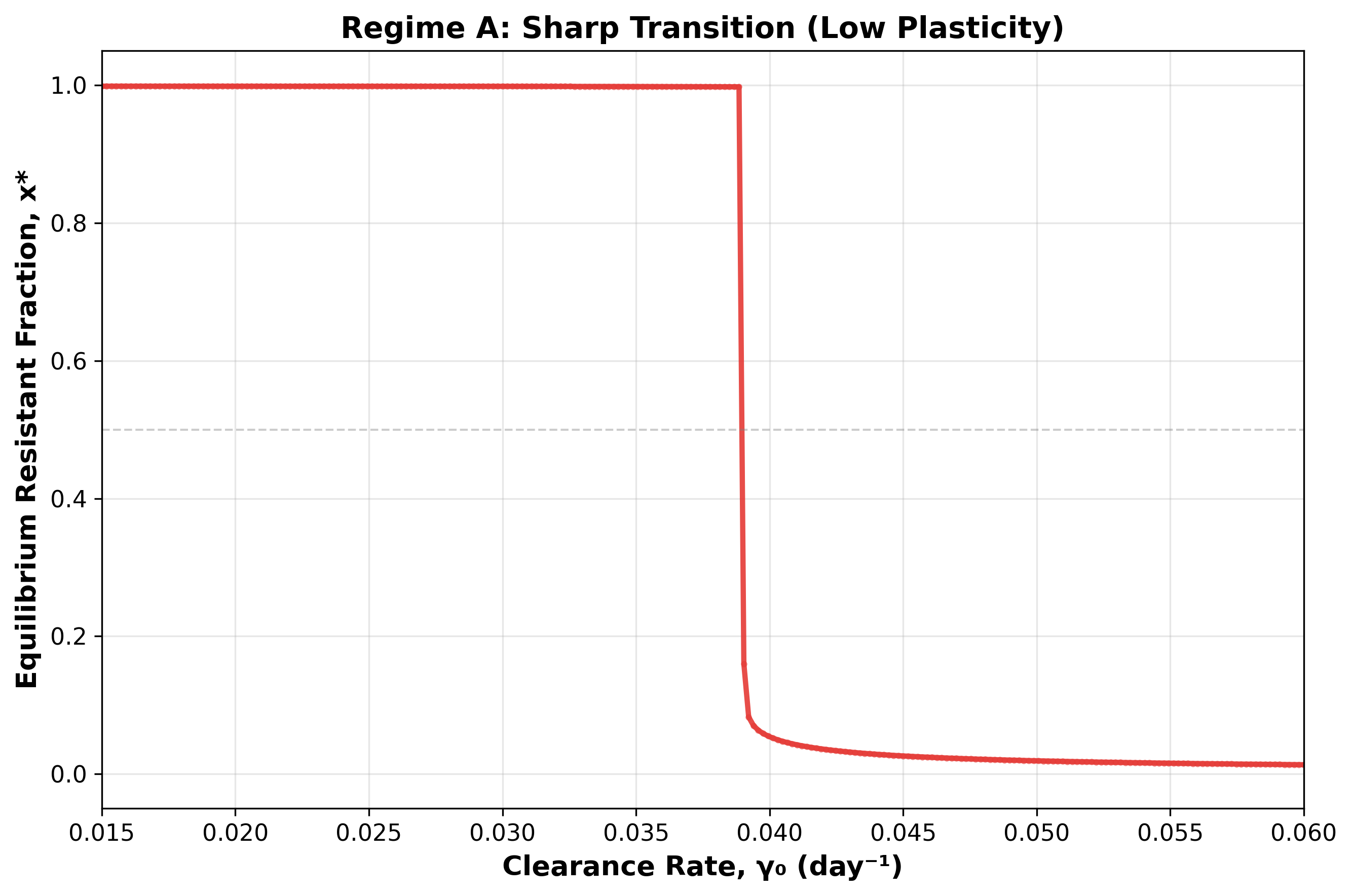}
\caption{\textbf{Regime A: Bifurcation with respect to clearance rate.} Equilibrium resistant-cell fraction, $x^*$, as a function of clearance rate, $\gamma_0$, at a fixed protection factor $\varphi = 0.70$. Each point represents the final state of a simulation integrated to equilibrium (2500 days). System exhibits sharp transition at $\gamma_0 \approx 0.039$ day$^{-1}$, separating high-resistance and low-resistance regimes.}
\label{fig:bifurcation_a_gamma}
\end{figure}

\subsubsection{Bifurcation with Respect to Protection Factor}

To examine whether acid-resistance machinery effectiveness exhibits similar threshold behavior, we computed equilibrium resistance across a range of protection factor values at three different clearance rates: below, at, and above the critical clearance threshold (Figure \ref{fig:bifurcation_a_phi}).

\textbf{Low clearance ($\gamma_0 = 0.025$ day$^{-1}$):}
At poor vascular perfusion, equilibrium resistance remains high ($x^* > 0.99$) across nearly the entire range of protection factors. Variations in $\varphi$ produce minimal changes in resistance, indicating that when acid clearance is insufficient, resistance evolution is inevitable regardless of acid-resistance machinery effectiveness. 

\textbf{Critical clearance ($\gamma_0 = 0.039$ day$^{-1}$):}
At the critical clearance rate, protection factor produces a sharp bifurcation. Above $\varphi \approx 0.58$, the system equilibrates to high resistance ($x^* > 0.95$), while below this threshold, resistance is almost suppressed ($x^* < 0.1$). This indicates that at marginal perfusion levels, acid-resistance machinery effectiveness becomes a critical determinant of evolutionary outcomes. 

\textbf{High clearance ($\gamma_0 = 0.052$ day$^{-1}$):}
At good vascular perfusion, equilibrium resistance remains low ($x^* < 0.01$) across the entire range of protection factors. As with poor perfusion, variations in $\varphi$ produce minimal effect, indicating that sufficient acid clearance can maintain sensitivity even in the presence of functional acid-resistance machinery.

\textbf{Interpretation:}
The context-dependent effect of protection factor reveals a critical parameter interaction in low-plasticity tumors. The effectiveness of targeting acid-resistance machinery depends strongly on baseline vascular perfusion: below a maximum unit clearance rate threshold, resistance is inevitable; above it, resistance is suppressed; only at intermediate clearance rates does $\varphi$ matter. This suggests a sequential therapeutic strategy: improve perfusion first to reach the critical clearance threshold, then inhibit acid-resistance machinery to tip the system toward sensitivity.

\begin{figure}[htbp]
\centering
\includegraphics[width=\textwidth]{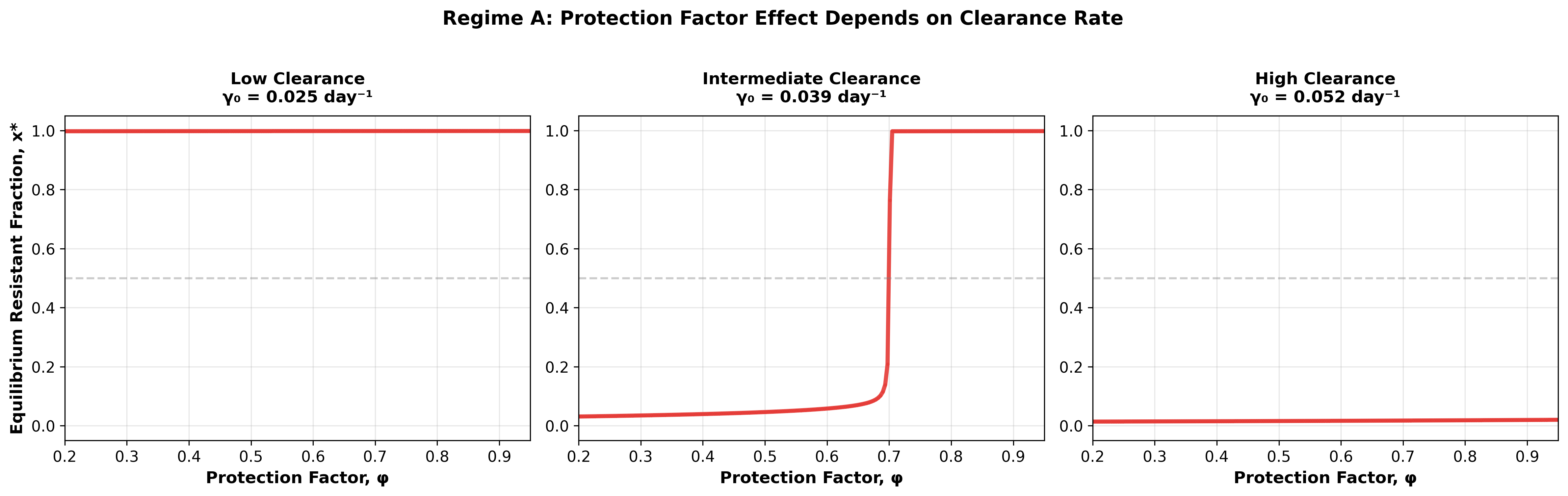}
\caption{\textbf{Regime A: Context-dependent bifurcation with respect to protection factor.} Equilibrium resistant fraction $x^*$ as a function of protection factor, $\varphi$, at three clearance rates: $\gamma_0 = 0.025$ day$^{-1}$ (below bifurcation threshold), $\gamma_0 = 0.039$ day$^{-1}$ (at bifurcation threshold), and $\gamma_0 = 0.052$ day$^{-1}$ (above bifurcation threshold). Protection factor produces a sharp bifurcation only at intermediate clearance rates, demonstrating parameter interaction. At low clearance, resistance is inevitable regardless of $\varphi$; at high clearance, sensitivity is maintained regardless of $\varphi$. This indicates that targeting acid-resistance machinery is only effective within a specific range of vascular perfusion levels.}
\label{fig:bifurcation_a_phi}
\end{figure}

\subsection{Regime B: Continuous Transitions}

\subsubsection{Time-Series Dynamics}

Regime B exhibits qualitatively different dynamics, showing gradual shifts in resistance as clearance varies, allowing for intermediate equilibrium values of resistance where there is significant coexistence of both acid-resistant and sensitive cell populations (Figure \ref{fig:regime_b}).

\textbf{Low clearance scenario ($\gamma_0 = 0.023$ day$^{-1}$):}
The tumor evolves to a moderately resistant state with $x = 0.523$. Tumor pH equilibrates at 6.96, representing moderate acidosis ($h \approx 2.5$). Density equilibrates to $n = 0.882$.

\textbf{High clearance scenario ($\gamma_0 = 0.032$ day$^{-1}$):}
With improved clearance, equilibrium resistance decreases to $x = 0.322$. Tumor pH equilibrates to a higher ph at 7.04 ($h \approx 2.1$). Density equilibrates to $n = 0.882$.

\begin{figure}[htbp]
\centering
\includegraphics[width=\textwidth]{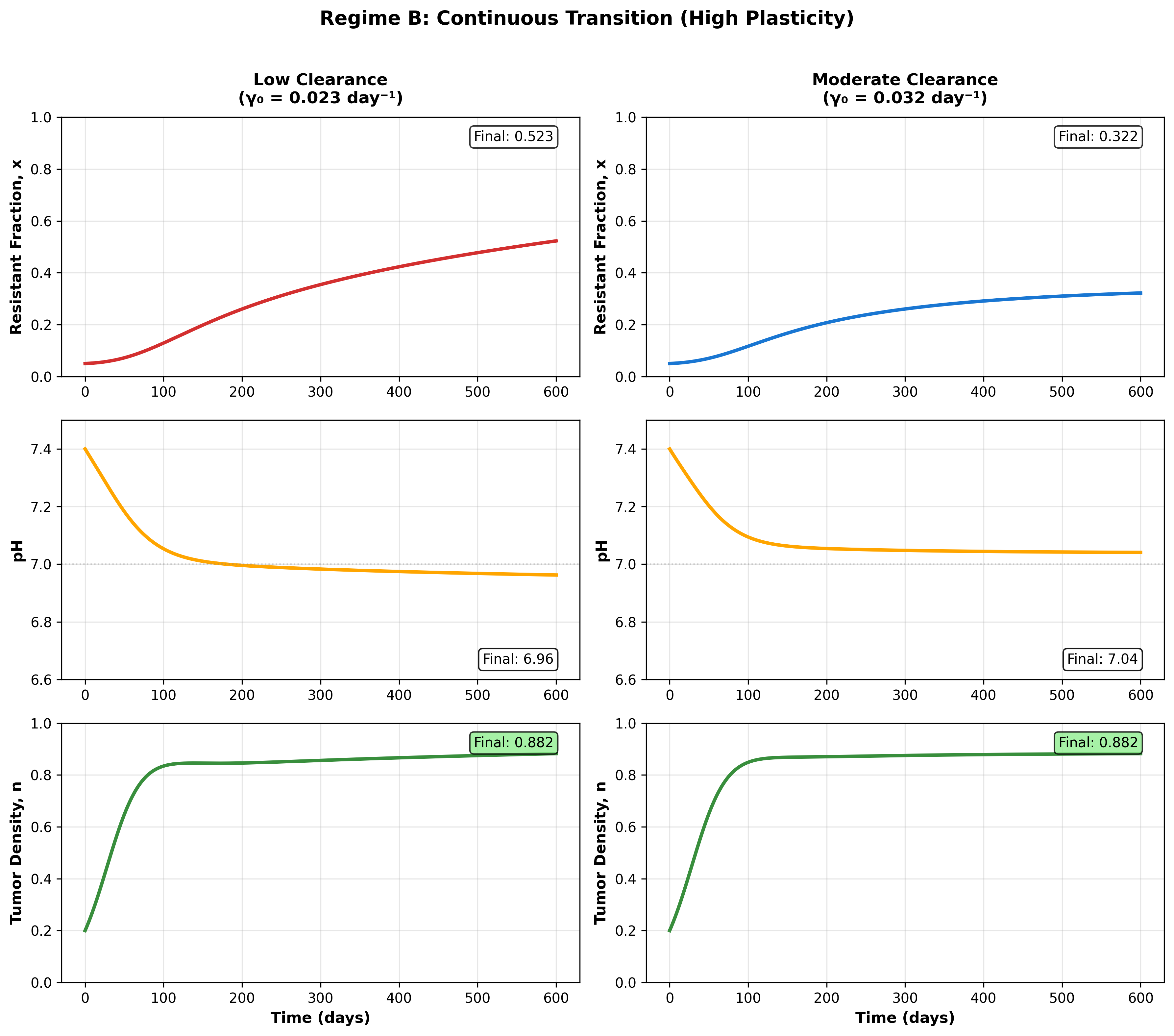}
\caption{\textbf{Regime B: Time-series dynamics in high-plasticity tumors.} Time evolution of resistant fraction $x(t)$, tumor pH, and tumor density $n(t)$ for low clearance ($\gamma_0 = 0.023$ day$^{-1}$, left) and moderate clearance ($\gamma_0 = 0.032$ day$^{-1}$, right). All other parameters identical. Initial conditions: $(x_0, h_0, n_0) = (0.02, 1.0, 0.15)$. Parameter values shown in table.}
\label{fig:regime_b}
\end{figure}

\subsubsection{Bifurcation with Respect to Clearance Rate}

The bifurcation diagram for Regime B (Figure \ref{fig:bifurcation_b_gamma}) shows markedly different behavior from Regime A. Rather than a sharp transition, equilibrium resistance varies smoothly and monotonically with clearance rate. As $\gamma_0$ increases from 0.018 to 0.038 day$^{-1}$, resistance decreases continuously with no threshold behavior. This indicates the absence of a bifurcation but rather a single stable equilibrium whose value depends smoothly on $\gamma_0$.

\begin{figure}[htbp]
\centering
\includegraphics[width=0.8\textwidth]{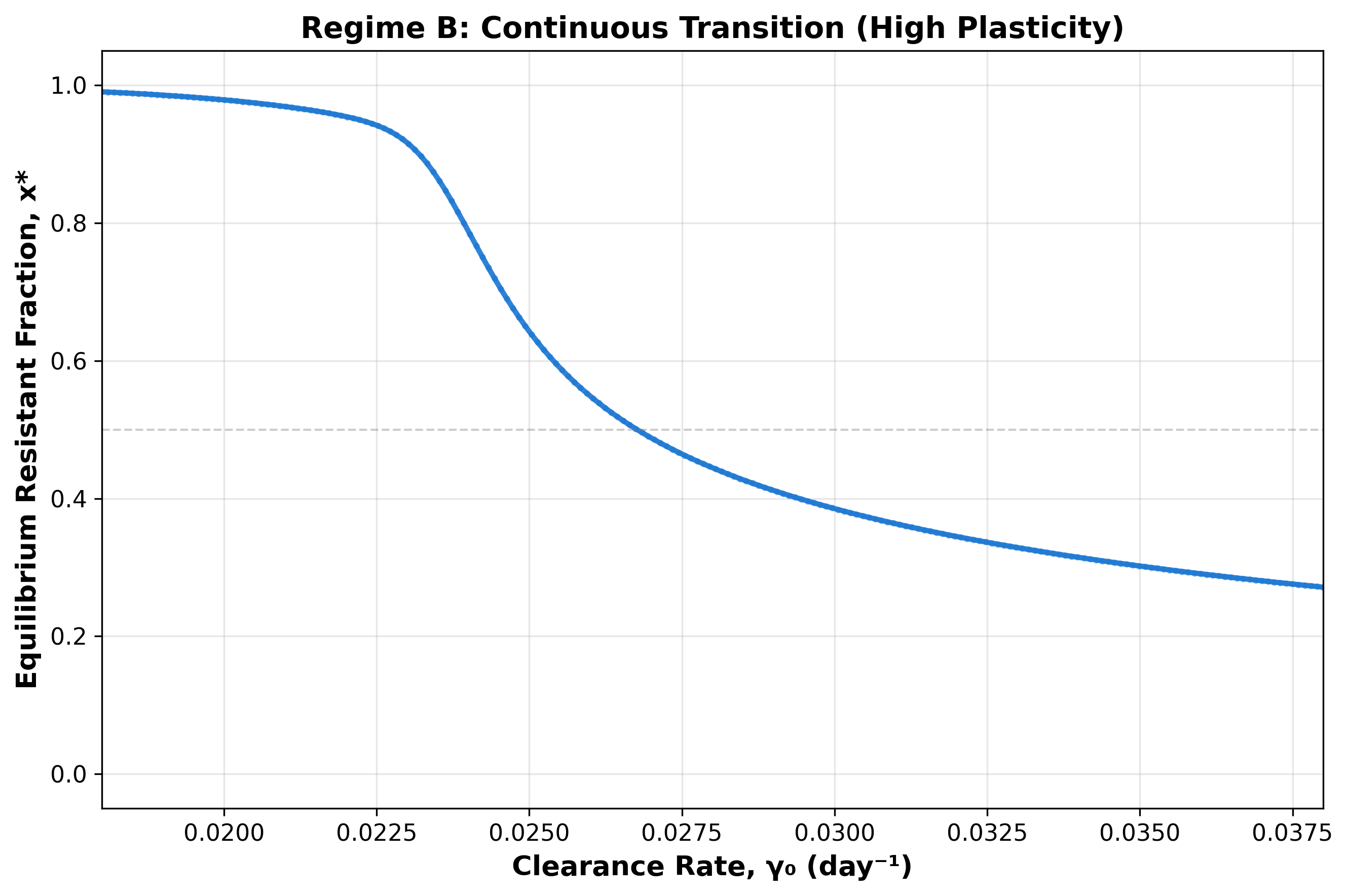}
\caption{\textbf{Regime B: Continuous transition with respect to clearance rate.} Equilibrium resistant-cell fraction $x^*$ as a function of clearance rate $\gamma_0$ at fixed protection factor $\varphi = 0.95$. System exhibits smooth, monotonic dependence on clearance with no threshold behavior. Resistance decreases continuously as clearance improves, indicating continuous adaptation to environmental conditions without catastrophic transitions.}
\label{fig:bifurcation_b_gamma}
\end{figure}

\subsubsection{Combined Effects of Clearance and Protection}

To characterize the joint effects of vascular perfusion and acid-resistance machinery in high-plasticity tumors, we computed equilibrium resistant-cell fraction across the two-dimensional parameter space spanned by maximal unit clearance rate, $\gamma_0$, and protection factor, $\varphi$ (Figure \ref{fig:bifurcation_b_phi}).

The resulting parameter landscape reveals smooth, continuous variation in equilibrium resistance across the entire domain. Resistance decreases monotonically with increasing clearance rate (vertical gradient) and decreasing protection factor (horizontal gradient). The approximately diagonal contour lines indicate that clearance and protection have independent, additive effects: reducing $\varphi$ by a given amount produces similar reductions in resistance regardless of $\gamma_0$, and vice versa.

Notably, no sharp boundaries or threshold behaviors emerge anywhere in the parameter space. The smooth gradients indicate that small improvements in either perfusion or machinery inhibition produce proportional reductions in resistance. This contrasts sharply with Regime A, where the same parameters exhibit context-dependent threshold effects.
\begin{figure}[htbp]
\centering
\includegraphics[width=0.7\textwidth]{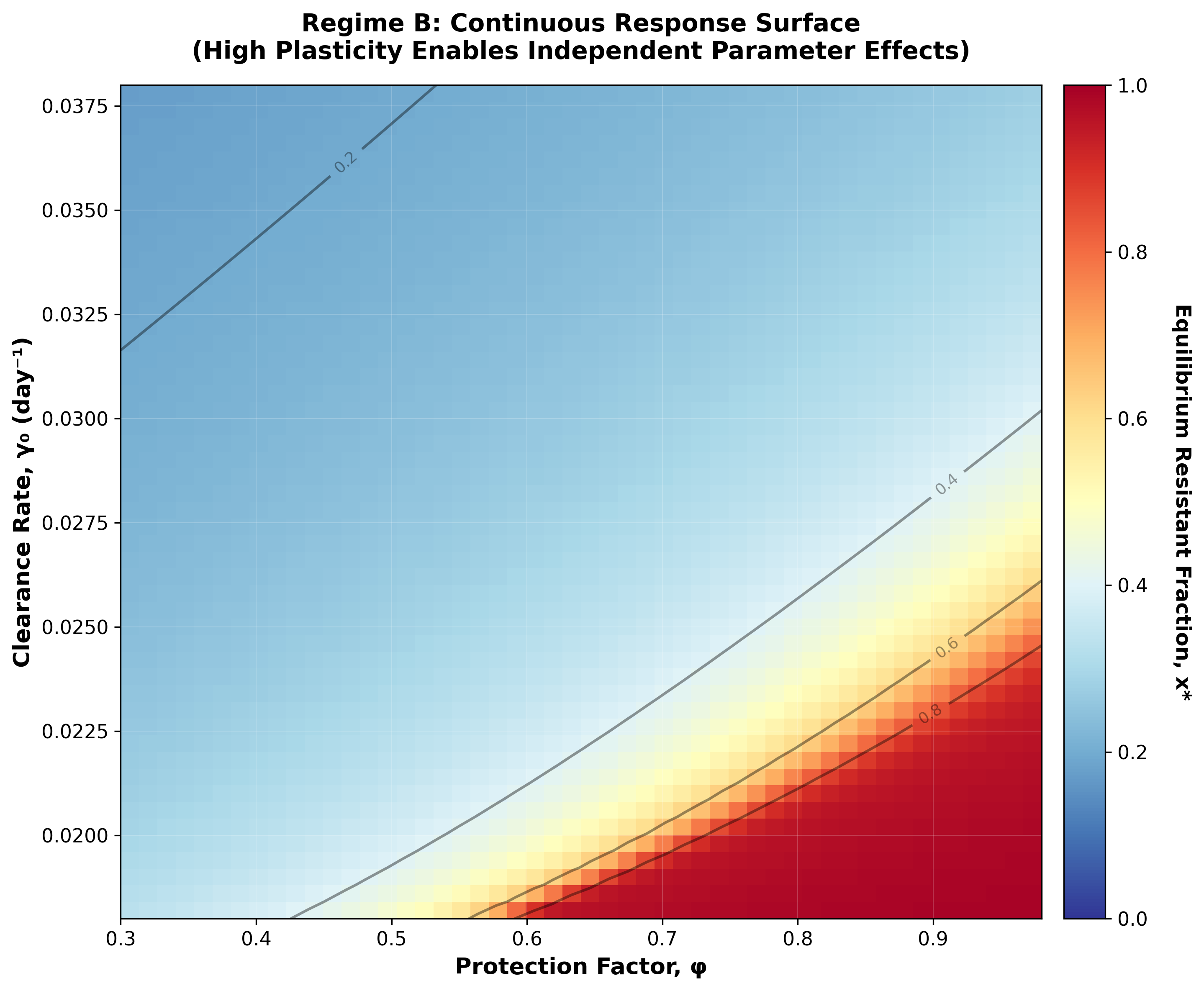}
\caption{\textbf{Regime B: Continuous response surface across parameter space.} Equilibrium resistant-cell fraction $x^*$ as a function of both clearance rate $\gamma_0$ (vertical axis) and protection factor $\varphi$ (horizontal axis). Color indicates equilibrium resistance level, with contour lines showing iso-resistance curves. The smooth, continuous gradients in both directions demonstrate that clearance and protection have independent, additive effects. No threshold behaviors or sharp boundaries emerge anywhere in the parameter space. This indicates that in high-plasticity tumors, incremental improvements in either vascular perfusion or acid-resistance machinery inhibition produce proportional reductions in resistance, suggesting combination therapy should produce additive benefits.}
\label{fig:bifurcation_b_phi}
\end{figure}

\textbf{Interpretation:}
The continuous response surface demonstrates that in high-plasticity tumors, clearance rate and protection factor act as independent control parameters with additive effects. This suggests a combination therapeutic strategy: simultaneously targeting vascular normalization and acid-resistance machinery inhibition should produce additive benefits, with each intervention contributing proportionally to resistance reduction regardless of the other's effectiveness.
\subsection{\ {Sensitivity Analysis}}

\ {Global sensitivity analysis identified the parameters with greatest influence on equilibrium resistance (Figure~\ref{fig:prcc}). The six most influential parameters, ranked by $|\text{PRCC}|$, were: the critical acidity threshold $h_c$ (PRCC $= -0.66$), stress-induced switching rate $\mu_0$ (PRCC $= +0.47$), vascular clearance rate $\gamma_0$ (PRCC $= -0.47$), metabolic cost $c$ (PRCC $= -0.45$), acid production rate $\alpha$ (PRCC $= +0.41$), and intrinsic growth rate $r$ (PRCC $= +0.40$). Two parameters showed no significant correlation with $x^*$: excess acid production by resistant cells $\beta$ ($p = 0.96$) and clearance reduction strength $\eta$ ($p = 0.15$), indicating that model predictions are robust to uncertainty in these quantities. The PRCC signs align with biological expectations: parameters that intensify microenvironmental acidosis ($\mu_0$, $\alpha$, $r$) or lower the threshold for resistance advantage (decreasing $h_c$) promote resistance evolution, while parameters that alleviate acidosis ($\gamma_0$) or impose fitness costs ($c$) suppress it. Notably, the protection factor $\varphi$ exhibited modest influence (PRCC $= +0.17$), consistent with our bifurcation analysis showing its effect is context-dependent---$\varphi$ matters primarily at intermediate clearance rates (Figure~\ref{fig:bifurcation_a_phi}). From a therapeutic standpoint, three of the six most influential parameters represent plausible intervention targets: the switching rate $\mu_0$, modifiable via hypomethylating agents; the clearance rate $\gamma_0$, improvable through vascular normalization; and, with weaker influence, the protection factor $\varphi$, targetable via acid-resistance machinery inhibitors. The remaining influential parameters ($h_c$, $c$, $\alpha$, $r$) represent intrinsic tumor characteristics that may serve as prognostic biomarkers but are not directly targetable.}

\begin{figure}[h]
\centering
\includegraphics[width=0.85\textwidth]{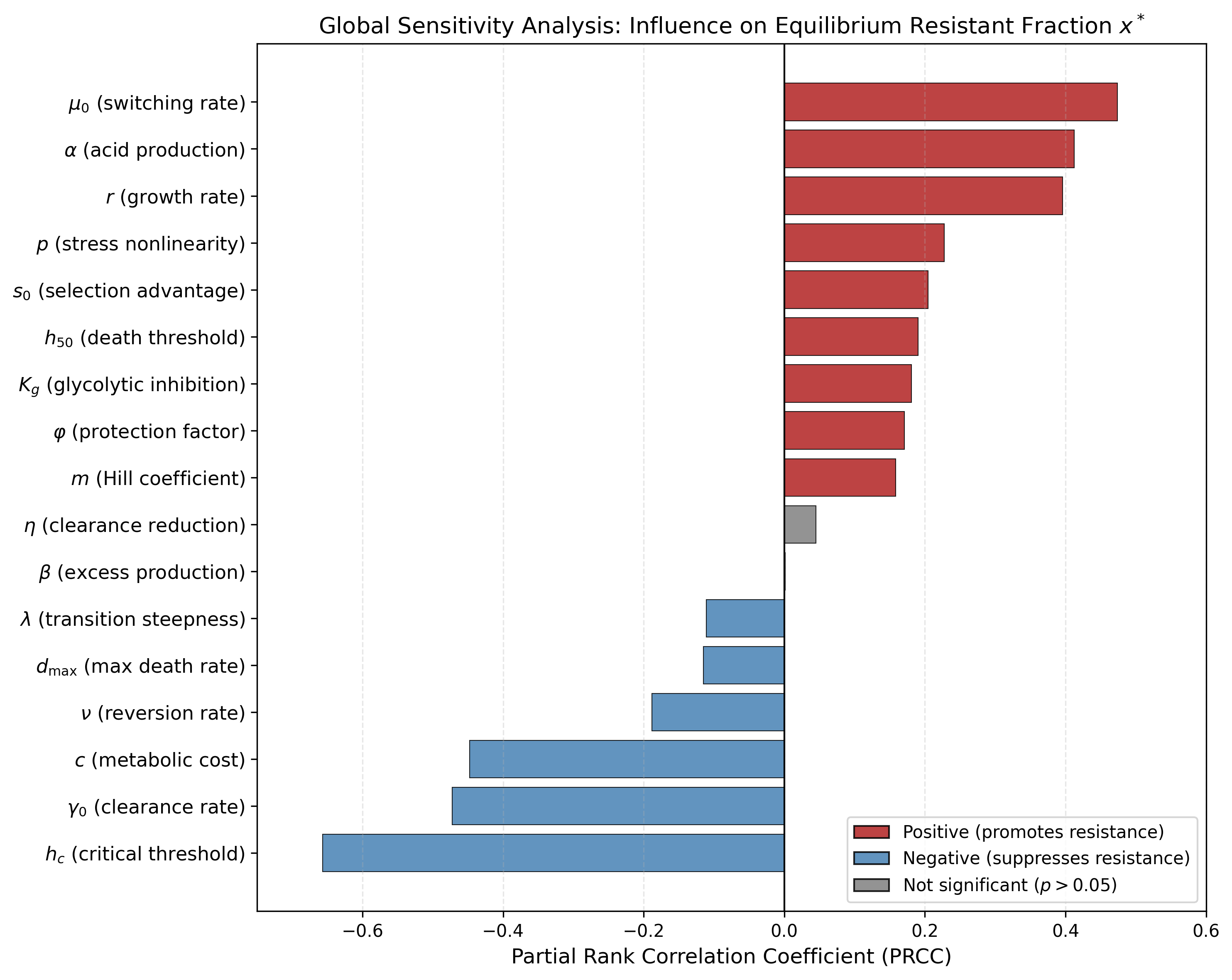}
\caption{\ {\textbf{Global sensitivity analysis of equilibrium resistance.} Partial Rank Correlation Coefficients (PRCC) quantifying parameter influence on equilibrium resistant fraction $x^*$, computed from $N = 1000$ Latin Hypercube samples. Positive values (red) indicate parameters whose increase promotes resistance; negative values (blue) indicate suppression of resistance; gray indicates no significant correlation ($p > 0.05$). The most influential parameters are $h_c$, $\mu_0$, $\gamma_0$, $c$, $\alpha$, and $r$. Model predictions are robust to $\beta$ and $\eta$.}}
\label{fig:prcc}
\end{figure}

\section{Discussion and Conclusions}
We proposed and analyzed a model for the development of resistance to acidosis in a vascularized tumor with fraction of resistant cells, normalized acidity, and total tumor density are used as state variables of a 3-dimensional dynamical system. We established the positive invariance and global existence of solutions and proved the existence of a unique extinction equilibrium that is locally asymtotically stable when the basic reproduction number of the population, $\mathcal{R}_0$ is below the threshold value of 1, and unstable if $\mathcal{R}_0>1$. We also proved that, when there is no phenotypic reversion, there exists a unique positive equilibrium with no sensitive cells, which we characterized explicitly.

Based on numerical simulations, we analyzed the evolution of resistance in the tumor under two different regimes: a low plasticity tumor --characterized by low stress-induced switching rate ($\mu_0=5\times 10^{-5}\ 
{\mathrm{day}}^{-1}$) and moderate protection ($\varphi = 0.70$), and a high plasticity tumor -- characterized by high stress-induced unit switching rate ($\mu_0 = 8 \times 10^{-4}\ 
{\mathrm{day}}^{-1}$) and strong protection ($\varphi = 0.95$), representing highly adaptive tumors with robust acid-resistance machinery.

For low plasticity tumors we found threshold-dependent dynamics, where a critical value $\gamma_0^{{\mathrm{crit}}}$ of $\gamma_0$ determines qualitatively distinct evolutionary outcomes. For $\gamma_0<\gamma_0^{{\mathrm{crit}}}$, the tumor evolves toward a high-resistance state, with tumor pH corresponding to severe acidosis, and a relatively high tumor density. The system reaches
a stable equilibrium characterized by a predominantly resistant population maintaining a chronically
acidic micro-environment. In contrast, for $\gamma_0>\gamma_0^{{\mathrm{crit}}}$, the tumor evolves toward a low-resistance state, with tumor pH corresponding to mild acidosis, and higher tumor density, closer to carrying capacity due to reduced acid stress. The equilibrium corresponds to a predominantly sensitive population in a nearly physiological micro-environment. There seems to be a sharp bifurcation from one equilibrium being attractive it becoming unstable while another one appears and inherits its stability. On the other hand, the effect of the protection factor, $\varphi$, on the evolution of acidity in the tumor is negligible when $\gamma_0$ is not too close to $\gamma_0^{{\mathrm{crit}}}$, and it becomes significant only when $\gamma_0$ is close to $\gamma_0^{{\mathrm{crit}}}$. 

The qualitative outcome is very different for high-plasticity tumors. Such tumors show gradual shifts in resistance as clearance
varies, allowing for intermediate equilibrium values of resistance where there is significant coexistence
of both acid-resistant and sensitive cell populations. Numerical simulations in that regime suggest that no bifurcation occurs here but rather the one positive equilibrium maintains stability as $\gamma_0$ sweeps its range, showing only gradual shifts in resistance as clearance
varies, allowing for intermediate equilibrium values of resistance where there is significant coexistence. A similar phenomenon is observed with respect to the protection factor, $\varphi$. 

These regime-dependent dynamics suggest that distinct therapeutic strategies should be used according to tumor plasticity levels. For low-plasticity tumors that exhibit bifurcation behavior, treatment should focus primarily on improving vascular clearance $\gamma_0$ to exceed the critical threshold $\gamma_0^{\text{crit}}$. Vascular normalization can be achieved through metronomic chemotherapy or controlled use of anti-angiogenic agents that improve perfusion rather than destroying vessels \cite{dvorak2009blood,jain2013vascular}. Physical interventions including mild hyperthermia, exercise, or hyperbaric oxygen also have potential in adequately improving tumor perfusion \cite{schadler2016exercise,moen2012hyperbaric}. Since the protection factor $\phi$ has negligible effect except when $\gamma_0$ is near the bifurcation threshold, which is an unlikely clinical scenario requiring precise parameter alignment, targeting acid-resistance machinery would generally be ineffective in this regime. However, if vascular function approaches but cannot exceed $\gamma_0^{\text{crit}}$, sequential therapy first improving perfusion then targeting $\phi$ through MCT1/4 inhibitors such as AZD3965 \cite{beloueche2020mct}, carbonic anhydrase IX inhibitors like SLC-0111 \cite{mcdonald2020caix}, or NHE1 inhibitors \cite{amith2017nhe1} may tip the system across the bifurcation. In contrast, high-plasticity tumors benefit from immediate combination therapy targeting both parameters, as our analysis demonstrates that incremental improvements in either vascular perfusion or acid-resistance inhibition produce proportional, additive reductions in resistance without threshold constraints. 

\ {Several limitations of the present framework warrant discussion. First, the spatially homogeneous assumption neglects pH gradients and localized selection pressures that arise in heterogeneous tumor microenvironments; spatial extensions using reaction-diffusion formulations could capture these dynamics \cite{gatenby1996reaction,mcgillen2013general}. Second, the binary phenotype classification simplifies a likely continuous spectrum of acid resistance; partial differential equation models structured by resistance level represent a natural generalization \cite{beerenwinkel2015cancer}. Third, the model omits immune cell populations, whose function is known to be impaired under acidic conditions \cite{pilon2016neutralization,boedtkjer2020acidic}; incorporating tumor-immune interactions could reveal additional therapeutic opportunities. Fourth, vasculature is treated as static, whereas dynamic angiogenesis and vascular remodeling influence both acid clearance and tumor growth \cite{dvorak2009blood,jain2013vascular}; coupling to angiogenesis models would enhance physiological realism. Finally, the absence of clinical or experimental validation limits predictive confidence; fitting to longitudinal measurements of tumor pH and resistance markers represents an essential next step \cite{anemone2019imaging}.
}

Our model's identification of plasticity as a critical determinant of treatment response emphasizes the need for {biomarkers distinguishing these tumor phenotypes. Potential markers of phenotypic plasticity include expression levels of epigenetic modifiers (EZH2, DNMT1), epithelial-mesenchymal transition markers, or functional assays measuring adaptation rates to environmental stress \cite{yuan2019phenotypic,meacham2013tumour,salgia2018plasticity}.} \ {methods to assess tumor phenotypic plasticity. Unlike conventional biomarkers that capture static tumor state at a single timepoint, plasticity is a dynamic property representing the rate at which cells switch phenotypes under stress. Estimating the switching rate $\mu_0$ would require functional assays that expose tumor cells to controlled acidic conditions and measure the rate of phenotypic adaptation over time, for instance by tracking the emergence of acid-resistant subpopulations via flow cytometry or live-cell imaging \cite{yuan2019phenotypic,meacham2013tumour}. Developing such assays for clinical use represents a necessary step toward implementing regime-dependent therapeutic strategies.} This framework provides a foundation for regime-dependent interventions where treatment strategies are tailored to the tumor's evolutionary capacity: perfusion-focused therapy for low-plasticity tumors versus combination approaches for high-plasticity tumors.

\section*{Use of AI tools declaration}
The authors declare they have not used Artificial Intelligence (AI) tools in the creation of this article.
\section*{Acknowledgments}

The authors had no sources of funding.

\section*{Conflict of interest}

The authors declare no conflict of interest.

\end{document}